\documentclass[aps,twocolumn, secnumarabic,nobalancelastpage,amsmath,amssymb,nofootinbib]{revtex4}

\usepackage{lgrind} 
\usepackage{chapterbib}
\usepackage{color}
\usepackage{graphics} 
\usepackage{graphicx} 
\usepackage{subfigure}
\usepackage{longtable} 
\usepackage{bm} 
\usepackage[colorlinks=true]{hyperref}

\newcommand{\BQIC}{Berkeley Center for Quantum Information and Computation, Berkeley, California 94720 USA}
\newcommand{\DeptPhys}{Department of Physics, University of California, Berkeley, California 94720 USA}
\newcommand{\DeptChem}{Department of Chemistry, University of California, Berkeley, California 94720 USA}
\newcommand{\SNL}{Department of Scalable and Secure Systems Research, Sandia National Laboratories, Livermore, California 94550 USA}

\newcommand{\etal}{\textit{et al.} }

\newcommand{\erf}[1]{Eqn.~(\ref{#1})}

\begin{document}

\title{Optimality of qubit purification protocols in the presence of imperfections}

\author {Hanhan Li$^{1,2}$}
\email {h\_li@berkeley.edu}
\author{Alireza Shabani$^{1,3}$}
\author{Mohan Sarovar$^{4}$}
\author{K. Birgitta Whaley$^{1,3}$}

\address{$^1$\BQIC}
\address{$^2$\DeptPhys}
\address{$^3$\DeptChem}
\address{$^4$\SNL}

\date{\today}

\begin{abstract}

Quantum control is an essential tool for the operation of quantum technologies such as quantum computers, simulators, and sensors. Although there are sophisticated theoretical tools for developing quantum control protocols, formulating \textit{optimal} protocols while incorporating experimental conditions remains a challenge. In this paper,
motivated by recent advances in realization of real-time feedback control in circuit quantum electrodynamics systems,
we study the effect of experimental imperfections on the optimality of qubit purification protocols.
Specifically, we find that the optimal control solutions in the presence of detector inefficiency and non-negligible
decoherence can be significantly different from the known solutions to idealized dynamical models.
In addition, we present a simplified form of the verification theorem to examine the global optimality of a control protocol.
\end{abstract}

\maketitle

\section{Introduction}
\label{sec:introduction}
Pure quantum states, states of systems with minimum classical uncertainty, are an ideal resource for many tasks in quantum information processing, including teleportation, quantum coding and error correction \cite{mikeandike}. However, frequently the states of systems encountered in the laboratory are mixed quantum states that contain classical uncertainty about various aspects of the particular physical system of interest. Ordinarily this uncertainty can be removed, and the state \textit{purified}, by an appropriate measurement or cooling procedure. Such purification is often a first, and critical, step in many quantum information processing, communication, and metrology protocols.

For many physical systems a measurement is properly treated as a finite timescale dynamical process as opposed to an instantaneous projective operation. In such systems, the measurement timescale (the time taken to complete a measurement and collect enough information to distinguish between the possible classical outcomes) is long enough that one can perform operations on the system during the measurement process. Such measurements are referred to as \textit{weak measurements}. Some examples of quantum information systems that can operate in regimes of weak measurement are quantum dots monitored by quantum point contacts \cite{Goan:2001gb}, and cavity QED implementations in optics \cite{Hoo.Lyn.etal-2000}, or the solid-state \cite{Vijay:2012}.

In the case of purification implemented by a weak measurement, it is natural to ask whether it is possible to accelerate the purification process by performing operations during the measurement. Jacobs showed in Ref. \cite{Jacobs:2003hc, Jacobs:2004tp} that in an ideal scenario it is possible to increase the instantaneous rate of purification by adding feedback operations that are unitary rotations conditioned on the information collected by the weak measurement thus far. Specifically, Jacobs showed that in the case of a single qubit with perfect efficiency measurement, no decoherence, and arbitrarily strong feedback (what we will call the ideal case), a feedback strategy that always maintains a two-level quantum system (qubit) in an unbiased basis with respect to the constant measurement basis results a maximization of the instantaneous rate of purification. We shall refer to this protocol as the \textit{unbiased measurement} protocol \footnote{It is implicit in the name that this protocol requires feedback to maintain the unbiased state.}. Subsequently, Refs. \cite{Combes:2006hx, Shabani:2008prl, Combes:2010gq} generalized this result and showed that in the ideal case it is possible to utilize feedback to increase the instantaneous rate of purification for arbitrary finite dimensional quantum systems. Wiseman and Ralph \cite{Wiseman:2006ti} have noted that it is useful to separate two different goals in the task of quantum state purification: the first goal, which we refer to as \texttt{max purity}, is that of maximizing the \textit{average purity} of the system at a \textit{given time}, while the second goal, which we refer to as \texttt{min time}, is that of minimizing the \textit{average time} taken to achieve a \textit{given purity}. These authors show that Jacobs' unbiased measurement strategy (which consequently maximizes the instantaneous rate of purification) is advantageous for the \texttt{max purity} goal, while a \textit{diagonal measurement} strategy, which measures in the diagonal basis of the qubit state (and requires no feedback), is better for the \texttt{min time} goal. In fact, Wiseman and Bouten \cite{Wiseman:2008bc} later proved that in the ideal case the unbiased measurement strategy is the optimal one for purifying qubits with the \texttt{max purity} goal and that the diagonal measurement strategy is optimal for purifying qubits with the \texttt{min time} goal. This highlights another reason why quantum state purification is an important problem in quantum control theory. It is one of very few problems in this domain where questions of optimality can be constructively addressed. In contrast to this situation for the state purification problem, the optimality of most quantum feedback protocols cannot be assessed in a constructive manner.

All the above works address the problem of quantum state purification in the ideal case where the measurements are of unit efficiency (i.e., where the measurement-induced state disturbance is equally compensated by a gain in information about the state \cite{Wiseman:2009vw}) and the feedback action is arbitrarily fast. Both idealizing assumptions must typically be relaxed in realistic systems. In Ref. \cite{Griffith:2007ct}, Griffth \etal relax the arbitrarily fast feedback assumption and consider the performance of both the unbiased measurement and diagonal measurement protocols for purifying the state of a superconducting Cooper pair box qubit. More recently, Combes and Wiseman \cite{Combes:2011wt} have analyzed the impact of a wider array of imperfections on the unbiased measurement protocol for purification, including finite strength feedback, time delay in the feedback loop, calibration errors, measurement inefficiency, and decoherence. Both of these studies indicate that the acceleration of purification rate by feedback is severely hampered by practical constraints.
 
In this work, we extend the study of quantum state purification by studying the optimality of purification protocols in the presence of key experimental imperfections. The imperfections we consider are measurement inefficiency and extrinsic decoherence as a result of environmental noise. Both these imperfections will be present in most quantum information processing architectures. Hence it is important to consider their effects on purification, and also to formulate optimal strategies for purification in their presence. We do not consider the imperfections arising from finite strength feedback or time delay in the feedback loop, since it is much more difficult to analyze optimality in the presence of these features. The remainder of the paper is structured as follows. In section 2 we introduce the physical system and dynamics we will analyze. In section 3, we find the time local optimal strategy for purification in the presence of imperfections. Section 4 discusses the global optimality of the locally optimal strategy for the \texttt{max purity} goal, using a simplified form of the verification theorem (as part of our calculations we derive a simplified form of the verification theorem \cite{Jacobs:2008vt} for verifying global optimality of control protocols, which is presented in appendix A).   We also compare the global optimality of the local strategy formulated here against other known protocols. Then the global optimality of local strategy for the \texttt{min time} goal is discussed in section 5.  A summary and conclusions are presented in section 6. 

\section{Dynamics}
\label{sec:dynamics}
We restrict our attention to the case of a quantum two-level system (qubit). Although this is the simplest finite dimensional system, it is also the most relevant from a quantum information perspective since physical implementations of qubits are the fundamental building blocks for most quantum information tasks. In addition, to examine the effect of imperfections on state purification, it suffices to examine the case of a qubit.

The system is subject to a weak, continuous measurement of an operator $M$  with strength $k$ and efficiency $\eta$. In addition, we assume the qubit is coupled to a low temperature environment which induces relaxation and decoherence dynamics on the qubit. The master equation describing the time evolution of a qubit with such dynamics is given by \cite{Bre.Pet-2002, Wiseman:2009vw}
\begin{align}
\label{eqa:evolution}
d\rho = & \gamma_{1}\mathcal{D}[\sigma_{-}]\rho dt+\gamma_{\phi}\mathcal{D}[\sigma_{z}]\rho/2 dt+\mathcal{D}[\sqrt{2k}M]\rho dt \nonumber\\
            &+ \mathcal{H}[\sqrt{2k\eta}M]\rho dW,
\end{align}
where $\rho$ is the qubit density matrix, $dW$ is a Wiener increment satisfying $dW^2=dt$, and we have set $\hbar=1$. The super-operators in this equation are defined as: $\mathcal{D}[A]\rho\equiv A\rho A^\dag -\frac{1}{2}(A^\dag A\rho+\rho A^\dag A)$ and $\mathcal{H}[A]\rho \equiv A\rho+\rho A^\dag-\mathrm{Tr}[(A+A^\dag)\rho]\rho$. This equation is in a rotating frame with respect to a free Hamiltonian of the form $\frac{1}{2}\omega(t)\sigma_z$, and $M$ and $\rho$ should be interpreted in this rotating frame. Here we have utilized the Born and Markov approximations of the noisy environment in order to summarize its effects on the qubit as Markovian dephasing at rate $\gamma_\phi$ and relaxation at rate $\gamma_1$. Since the environment is considered to be at low temperature ($k_B T \ll \omega$), we only consider its de-excitation (relaxation) effects on the qubit. The time-dependent measurement results, or measurement current, that generates conditioned evolution by \erf{eqa:evolution} can be expressed as:
\begin{equation}
    \label{eqa:current}
    I(t) =\sqrt{k} \mathrm{Tr}[(M+M^\dag)\rho(t)] +\xi(t)/\sqrt{\eta}.
\end{equation}
where  $\xi(t)=dW/dt$ in a white noise process. In this paper, we assume that the measurement is along the computational basis axis, i.e. $M=J_{z}=\sigma_{z}/2$ (note that this $M$ has no time dependence in the rotating frame).

Finally, we add a time-dependent coherent rotation of the qubit, $F(t)$, that constitutes our feedback Hamiltonian.  $F(t)$ could be based on the measurement results up to $t$ and generates the following dynamics in addition to \erf{eqa:evolution} \cite{Wiseman:2009vw}:
\begin{equation}
	\label{eqa:feedback}
	[\dot{\rho}]_{fb} = -\mathrm{i}[F(t),\rho].
\end{equation}

We consider feedback of arbitrary strength for convenience, including infinite strength feedback which is modeled as instantaneous unitary rotations at any time superposed on the evolution given by \erf{eqa:evolution} \footnote{Protocols that require ``infinite" feedback strength may be approximated reasonably well in circuit QED  \cite{Vijay:2012}, where microwave rotations are significantly faster than other relevant timescales, namely, $|F(t)| \gg k, \gamma_1, \gamma_\phi$.}. 

Because of rotational invariance about the $z$ axis, going away from the $x-z$ plane does not aid purification. Thus, without loss of generality, we may restrict our attention to the Bloch vector components $x$ and $z$, and consider the feedback rotation to be about the $y$ axis.

Without feedback, the evolution of the Bloch vector components of the qubit ($\rho = \frac{1}{2}(\mathbf{1} + x\sigma_x + y\sigma_y + z\sigma_z)$) is:
\begin{subequations}
    \label{eqa:components}
    \begin{align}
        dx&=-(\gamma_2+k)x dt-\sqrt{2k\eta}xz dW, \\
        dz&=-(\gamma_1+\gamma_1z)dt+\sqrt{2k\eta}(1-z^2)dW,
    \end{align}
\end{subequations}
where $\gamma_2=\gamma_1/2+\gamma_{\phi}$ and $r=\sqrt{x^2+z^2}$. 

Using Ito's lemma \cite{Wiseman:2009vw}, we can translate \erf{eqa:components} into the following dynamic equation for the variable $r=\sqrt{x^2+z^2}$, the length of the Bloch vector:
\begin{align}
	\label{eqa:dynamics}
	dr=&[r(\gamma_2-\gamma_1)+k(r-\frac{\eta}{r})]u^2 dt-\gamma_1 u dt \nonumber\\
	     & +[k(\frac{\eta}{r}-r)-\gamma_2 r]dt+\sqrt{2k\eta}(1-r^2)u dW,
\end{align}
with $u =\frac{z}{r}$. 
\footnote{The apparent singularity at $r=0$ can be removed by changing the state variable from $r$ to the purity, $P$, as we will see in \erf{eqa:puri_rate}. 
However, because the qubit dynamics are more readily visualized in terms of the Bloch vector $r$, and expressions look simpler in $r$, we will use $r$ as our state variable for the greater part of the presentation in this paper (in situations where the singularity does not affect our results). Numerical calculations of \erf{eqa:dynamics} are handled by setting $u=-1$ at the origin to avoid the singularity.}
Note that for $u=\pm 1$ we have $z = \pm r$ and the state lies on the $z$-axis. For $-1<u<1$, the Bloch vector makes a non-zero angle with the $z$ axis.
The feedback control (which is a unitary rotation) does not affect the above dynamical equation for $r$, but it does affect the dynamical equation for $u$ (which is not shown here). However, the assumption of arbitrary strength feedback simplifies the treatment since it implies that we can set $u$ arbitrarily by instantaneous rotation at any time. Therefore, we identify $r$ as our state variable and $u(r,t)\in[-1, 1]$ as our control input in the above dynamics. Some of the actual controls we consider for various optimality conditions below require infinite strength feedback while others do not, and we will make this requirement explicit when relevant.

Finally, we also write down the special case of \erf{eqa:dynamics} with no decoherence for later convenience. That is, when $\gamma_1=\gamma_\phi=0$,
\begin{equation}
    \label{eqa:dynamics_r}
    dr=k(r-\frac{\eta}{r})(u^2-1)dt+\sqrt{2k\eta}(1-r^2)u dW.
\end{equation}

\section{The Locally Optimal Strategy}
\label{sec:TheLocallyOptimalStrategy}
In this section, we will formulate a locally optimal strategy that maximizes the instantaneous rate of purification in the presence of measurement inefficiency and decoherence. 
This strategy is the generalization of the unbiased measurement 
protocol \cite{Jacobs:2003hc, Jacobs:2004tp} which is also a locally optimal strategy, but was also shown to be globally optimal for purifying qubits with the \texttt{max purity} goal, in the ideal case with no imperfections \cite{Wiseman:2008bc}. In the following sections, we will analyze whether this locally optimal strategy is also globally optimal for any goal in the presence of imperfections. 

We begin by writing an equation of motion for the purity, which is defined as: $P=\mathrm{Tr}[\rho^2]=\frac{1}{2}(1+r^2)$. Using Ito's Lemma and \erf{eqa:dynamics}, we obtain the rate of change of purity as:
\begin{align}
    \label{eqa:puri_rate}
    d P =& [\gamma_2-\gamma_1+k(1-2\eta+{\eta}r^2)]r^2 u^2 dt - \gamma_1 r u dt \nonumber\\
	& +[k\eta-(\gamma_2+k)r^2]dt +\sqrt{2k\eta}(1-r^2)ru dW.
\end{align}
This equation consists of a deterministic quantity and a stochastic quantity that is proportional to $dW(t)$. The former gives the rate of change of \textit{average} purity,  $\langle \dot{P}\rangle$, since $dW(t)$ averages to zero. Here the angle brackets indicates an average over the stochastic noise processes. Notice if write $r$ in terms of $P$ on the right hand side, $r=\sqrt{2P-1}$, this equation can be regarded as the dynamics for the state variable $P$. Maximizing the instantaneous \textit{average} purification rate, $\langle \dot{P}\rangle$ by choice of rotations around the $y$ axis is equivalent to maximizing the following quadratic function of $u$:
\begin{equation}
\label{eqn:fu}
    f(u)=[\gamma_2-\gamma_1+k(1-2\eta+{\eta}r^2)]r^2 u^2-\gamma_1 r u.
\end{equation}
The control that maximizes this function can be considered a \textit{locally} optimal strategy since it maximizes the instantaneous rate of change in average purity, and we label it $u_{lo}(t)$. It is not \textit{a priori} clear that such a locally optimal strategy will be globally optimal for either the \texttt{max purity} or \texttt{min time} purification goals, and we will investigate this issue in sections 4 and 5 below. 

\subsection{Local optimality in the absence of decoherence}
\label{sec:local_nodecoh}
Consider $\gamma_1=\gamma_\phi=0$, in which case, $f(u) \rightarrow f_{\textrm{no-decoherence}}(u) = k(1-2\eta+{\eta}r^2)r^2 u^2$. The maximizers of this function are easily found and are summarized in Table \ref{tab:local_optimal_r}. Interestingly, the $1/2<\eta< 1$ case introduces a fragmentation of the locally optimal control strategy that is not present in the ideal case ($\eta=1$) \footnote{We note that Combes and Wiseman have previously suggested that such a fragmented, or switching, strategy might be optimal in their study of purification under imperfections \cite{Combes:2011wt}.}. Also, Table \ref{tab:local_optimal_r} shows that if $\eta \leq 1/2$, the locally optimal strategy is simply to measure diagonally by keeping the state of the qubit in the $\sigma_z$ basis, i.e. $|u(t)|=1$. The simplest way to implement this \textit{diagonal measurement} strategy is to perform an initial instantaneous rotation to the $z$ axis (since control is assumed to be instantaneous and at no cost) and no successive feedback, and we will call this the \textit{no-feedback diagonal measurement} protocol. On the other hand, when $\eta=1$, which is the ideal case that was analyzed in Ref. \cite{Jacobs:2003hc} , we recover the unbiased measurement protocol as the locally optimal strategy: $u(t)=0$. This protocol keeps the qubit in an unbiased basis with respect to the measurement, and $u(t)=0 \implies z(t)=0$ is maintained by strong rotations.

In the intermediate case where $1/2 < \eta < 1$, a critical Bloch vector length emerges, $r^*=\sqrt{2-1/\eta}$, around which the locally optimal strategy switches between the diagonal measurement protocol and the unbiased measurement protocol. That is, when $r<r^*$ the feedback induced control $u=0$ maximizes the rate of change of average purity, while when $r \geq r^*$, the strategy of diagonal measurement maximizes this quantity. We note that there is a critical purity corresponding to the critical Bloch vector length, given simply by: $P^* = \frac{1}{2}(1+r^{*2}) = \frac{1}{2}(3-1/\eta)$.

A note is in order about the feedback nature of this locally optimal strategy. Obviously $|u(t)|=1$ requires no feedback since this corresponds to constant rate measurement along a fixed axis ($\sigma_z$) and $F(t)=0$. $u(t)=0$ on the other hand requires maintaining the Bloch vector along the $x$-axis despite measurement-induced fluctuations causing deviations from this axis. To do this, as specified in Jacobs' original unbiased measurement protocol  \cite{Jacobs:2003hc, Jacobs:2004tp, Combes:2011wt} the feedback Hamiltonian must be proportional to the measurement current: $F(t)=\sqrt{2k}\eta \frac{I(t)}{x(t)} J_{y}$. This is a conditioned rotation since it is inversely proportional to $x(t)$, the $x$ projection of the Bloch vector at the current time instant \footnote{Technically, this $F(t)$ is an unbounded Hamiltonian since $dW(t)$ is unbounded, and furthermore, the initial state $x(0)=0$. However, it has been shown that tempered approximations of this Hamiltonian suffice to implement the unbiased measurement protocol \cite{Combes:2011wt}.}. However, it should be noted that this protocol does not require real-time state estimation to execute. This is because in the presence of the feedback, when $u(t)=0$, the evolution of the $x$ component of the Bloch vector is deterministic, since the feedback effectively cancels the stochastic component of the evolution. In contrast, the locally optimal strategy when $1/2 < \eta < 1$ requires a switch between the diagonal measurement protocol and the unbiased measurement protocol when $r$ crosses $r^*$. In order to implement this, one requires a real-time estimate of the length of the Bloch vector, $r(t)$, which does not evolve deterministically when $u\neq 0$ as can be seen from \erf{eqa:dynamics}. Furthermore, this optimal strategy requires one to rotate the state between the $x$ and $z$ axes as the Bloch vector length crosses $r^*$. More precisely, for $r<r^*$, the feedback prescribed by the unbiased measurement protocol maintains the state on the $x$-axis (which is unbiased with respect to the measurement along $z$-axis). When the Bloch vector length increases to $r>r^*$, the locally optimal strategy prescribes a fast rotation of the state from the $x$-axis to the $z$-axis (a $\pi/2$ $\sigma_y$ rotation) followed by no feedback (unless $r<r^*$ again at a late time due to the stochastic evolution of purity under diagonal measurement measurement). These operations require real-time state estimation in addition to rotations that take negligible time. Such requirements make implementation of the locally optimal strategy challenging when $1/2 < \eta < 1$. 

\begin{table}[h]
\caption{The strategy that maximizes the instantaneous rate of increase of average purity (the locally optimal strategy) when $\gamma_1=\gamma_\phi=0$. $r^*=\sqrt{2-1/\eta}$ is the critical Bloch vector length at which there is a discontinuous change in protocols when $1/2<\eta <1$.}
\label{tab:local_optimal_r}
\begin{ruledtabular}
\begin{tabular}{cccc}
 & $\eta\leq 1/2$ & $1/2<\eta <1$ & $\eta=1$\\
\hline
 $u_{lo}(r,t)$ & $\pm 1$ & $\left\{
 \begin{array}{cl}
		0 & , \; r\leq r^* \\
		\pm 1 & , \; r> r^* \\
 \end{array}
 \right.$
 & $0$
\end{tabular}
\end{ruledtabular}
\end{table}

\begin{figure}[h]
    \centering
        \includegraphics[width=0.45\textwidth]{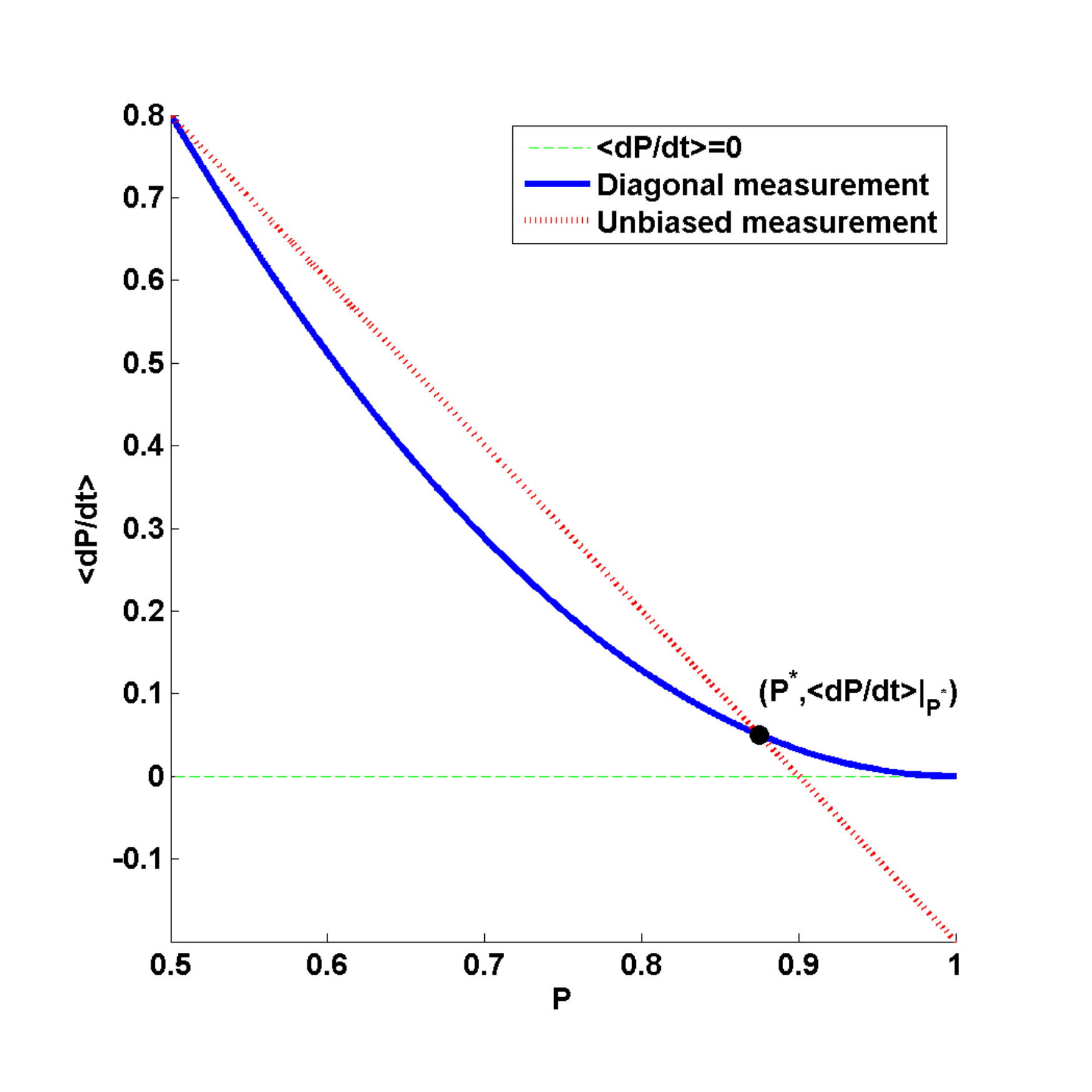}
    \caption{Average rate of change of purity as a function of instantaneous purity in the presence of measurement inefficiency, for the diagonal measurement protocol ($|u(r,t)| = 1$) and the unbiased measurement protocol ($u(r,t) = 0$). The point where the locally optimal strategy switched between these protocols is indicated as $P^*$. Parameters used for this plot are $k=1$, $\eta=0.8$, $\gamma_1=\gamma_\phi=0$. }
   \label{fig:dpurity_vs_purity}
\end{figure}

To illustrate the behavior of purification in the case when $1/2<\eta <1$, Figure \ref{fig:dpurity_vs_purity} shows the average rate of change of purity as a function of the instantaneous purity under both the unbiased measurement protocol and the diagonal measurement protocol. The rate of purification decreases with the instantaneous purity for both protocols, but while this rate is always positive for the diagonal measurement protocol, it can be negative for the unbiased measurement protocol when the purity is large. This is the reason it is advantageous to switch to the diagonal measurement protocol at large purity values. Physically, the reason for this switch is that the feedback required for the unbiased measurement protocol is non-ideal for inefficient measurement, and hence for Bloch vectors that are already large it is preferential to switch off the non-ideal feedback.

\subsubsection{Analytical solution for purity evolution when $1/2 < \eta < 1$}
\label{sec:analytical_soln}
When $\eta < 1/2$, the locally optimal strategy is simply measurement along a fixed basis, and in this case an analytical form for the probability distribution for $r$, the length of the Bloch vector, is easily computed and given in section 4.1 below (since the evolution is simply diffusion along the $z$-axis after a possible initial rotation to move the state to this axis). Similarly, when $\eta=1$, when the locally optimal strategy is the unbiased measurement protocol, an expression for the probability distribution for $r$ is given in Ref. \cite{Jacobs:2003hc} (not in closed form however). Here we complete this characterization and calculate an analytical expression for the probability evolution for $r$ in the case of qubit evolution under the locally optimal strategy when $1/2<\eta < 1$. Note that this case is significantly more complicated than the other two mentioned above since it involves a switching of protocols around the point $r^*$.

Consider a known initial state of the qubit on the $z$-axis with $r(0) = r_0$ (\textit{e.g.} $r_0=0$ when $\rho_0 = \mathbf{I}/2$). The probability distribution function $p(r,t)$ of the Bloch vector length at time $t$ is given by the following Fokker-Planck equation with the initial condition $p(r,0)=\delta(r-r_0)$:
\begin{align}
    \label{eqa:fokkerplanck}
    \frac{\partial}{\partial t}p(r,t)=& -\frac{\partial}{\partial r}[k(\frac{\eta}{r}-r) \mathrm{\theta}(r^*-r) p(r,t)]  \nonumber\\
    &+\frac{\partial^2}{\partial r^2}[k\eta(1-r^2)^2 \mathrm{\theta}(r-r^*) p(r,t)],
\end{align}
where $\theta(x)$ is the Heaviside step function.

First consider $r_0\geq r^*$. In this case the locally optimal strategy implements the diagonal measurement unless the random (diffusive) evolution results in $r(t)\leq r^*$ at some future time $t$. But when this happens the strategy switches to the unbiased measurement protocol which deterministically increases purity until $r(t+\Delta t)>r^*$ and we are returned to the region where the diagonal measurement protocol is preferred. Therefore, for this initial condition $r$ will not go below $r^*$ for a finite time. Hence we consider the following ansatz for the probability distribution function for $r$:
\begin{equation}
    \label{eqa:distribution}
    p(r,t)=p_1(t)\delta(r-r^*)+p_2(r,t) \theta(r-r^*).
\end{equation}
Given this ansatz, the Fokker-Planck equation can be translated into the following set of equations with initial conditions $p_2(r,0)=\delta(r-r_0)$ and $p_1(0)=0$:
\begin{subequations}
    \label{eqa:pde}
    \begin{align}
        & r\eta(1-r^2)^2 p_2(r^*,t)=(\eta-r^2)p_1(t),\\
        & \frac{\partial}{\partial r}[k\eta (1-r^2)^2 p_2(r,t)]|_{r^*}=\frac{\partial}{\partial t}p_1(t),\\
        & \frac{\partial}{\partial t} p_2(r,t)=\frac{\partial^2}{\partial r^2}[k\eta(1-r^2)^2 p_2(r,t)].
    \end{align}
\end{subequations}
Note that in the region $r>r^*$ \erf{eqa:pde}c is the Fokker-Planck equation. \erf{eqa:pde}b enforces probability conservation at the boundary $r=r^*$. This set of equations can be solved by an appropriate change of variables: we calculate the solution explicitly in Appendix \ref{sec:solvingpde}. This analytical solution will be utilized when assessing global optimality in the section below. 

Now consider the alternative initial condition where $r_0<r^*$. In this case, the locally optimal strategy implements the unbiased measurement protocol,which will deterministically increase the purity of the qubit according to:
\begin{equation}
    \label{eqa:drift}
    r(t)=\sqrt{\eta-(\eta-r_0^2)\mathrm{e}^{-2kt}}
\end{equation}
This continues until the critical Bloch vector length is reached and the locally optimal strategy then switches over to the diagonal measurement protocol. Let $r(t^*)=r^*$. Then, after $t^*$, the non-deterministic diagonal measurement protocol is employed and we simply solve for the qubit dynamics that this prescribes, i.e., solve \erf{eqa:pde} for the distribution function with the initial condition $p_2(r,t^*)=\delta(r-r^*)$.

Together with the explicit characterizations of evolution under the locally optimal strategies when $\eta < 1/2$ and $\eta=1$ described above, this analytic characterization of the probability distribution for the Bloch vector length 
now provides a complete picture of the evolution of qubit purity under the locally optimal strategy for any measurement efficiency. 

\subsection{Local optimality in the presence of decoherence}
\label{sec:local_decoh}
In this subsection, we expand our analysis of locally optimal strategies that maximize the instantaneous rate of increase of average purity to the case where decoherence is present. Recall that the general expression for instantaneous average purification rate, which we want to maximize, is given in \erf{eqn:fu}. For general $\gamma_1$, $\gamma_\phi$, and $\eta$, in order to maximize $f(u)$ we require:
\begin{displaymath} u_{lo}(r,t) = 
\left\{ \begin{array}{l}
-1 ~ \textrm{if}~ \gamma_1+2r[\gamma_2-\gamma_1+k(1-2\eta+\eta r^2)]\geq 0 \\ 
\frac{\gamma_1}{2r[\gamma_2-\gamma_1+k(1-2\eta+\eta r^2)]} ~~~ \textrm{otherwise}
\end{array} \right. 
\end{displaymath}

For the explicit calculations in this work, we will focus on the most realistic situation, where $\gamma_1\neq 0$, $\gamma_\phi\neq 0$, and $\eta<1$.  

Here, the counterpart of the regime where $\eta\leq 1/2$ in the previous case of no decoherence is the parameter space that yields $u_{lo}(r,t)\equiv -1$. A sufficient condition for this is $\gamma_\phi+k(1-2\eta)\geq 0$. In the presence of decoherence, the term $-\gamma_1 ru$ term in $f(u)$ (see \erf{eqn:fu})  ensures that $u=+1$ is no longer a maximizing control. Physically, this simply means that because of uncontrollable relaxation (a $T_1$ process), it is not advantageous to attempt to purify to the excited state of the qubit. Instead, maintaining $u=-1 \implies z<0$ is the locally optimal strategy. Now to keep the $z$-component of the qubit negative, we must flip the qubit (e.g., apply $\pi$ rotation around $y$ axis) whenever our $\sigma_z$ measurement indicates that it is positive. Therefore this control corresponds to simply having a constant $\sigma_z$ measurement (which induces diffusive motion along the $z$ axis) that is interrupted by strong $\sigma_y$ rotations whenever $z(t)>0$. Such a feedback protocol requires continuous real-time state estimation. We shall refer to this strategy of maintaining the $z$-component of the qubit negative as the \textit{negative diagonal measurement} protocol.

The negative diagonal measurement protocol is expected to have better performance than the no-feedback diagonal measurement protocol\footnote{While the no-feedback diagonal measurement may come with an initial rotation to either $+z$ or $-z$ axis in the decoherence free regime, we will choose the more advantageous initial rotation to the $-z$ axis here.} both locally and globally, as long as $\gamma_1>0$. The $\sigma_z$ measurement has a chance of moving the qubit to the $z>0$ region and when this happens,  the strong feedback rotation in the negative diagonal measurement protocol will change the term $-\gamma_1 z$ in \erf{eqa:puri_rate} to $+\gamma_1 z$, while keeping all others invariant.  This will generate a larger instantaneous purification rate. Therefore, for any noise realization $dW(t)$, the negative diagonal measurement protocol gives a trajectory that has a purity larger than or equal to evolution under the no-feedback diagonal protocol at any time. We will see this manifest in the numerical simulations assessing global optimality in sections \ref{sec:max_purity} and \ref{sec:min_time}.

In the parameter space where $u_{lo}(r, t)\not\equiv 1$, $u_{lo}(r, t)$ has complicated dependence on $r$.  We have plotted its value for a typical set of parameters in Fig. \ref{fig:noisy_u_lo}.  When the condition $\gamma_1+2r[\gamma_2-\gamma_1+k(1-2\eta+\eta r^2)]\geq 0$ is not met, one must perform precise rotations around the $y$ axis so that $u = \frac{\gamma_1}{2r[\gamma_2-\gamma_1+k(1-2\eta+\eta r^2)]}$ is maintained. This also requires continuous real-time state estimation, and furthermore, requires precise knowledge of all the parameters in the system. Executing this locally optimal strategy in the presence of decoherence is hence very challenging from a practical standpoint.

The locally optimal strategy in the presence of decoherence is significantly more complex than that in the absence of decoherence. As a result we have been unable to formulate an analytical solution for the probability distribution of the Bloch vector length in this case.

\section{Global Optimality for the \lowercase{\texttt{max purity}} goal}
\label{sec:max_purity}
In this section, we will consider the \texttt{max purity} purification goal, i.e., to maximize $P(t)= \frac{1}{2} + \frac{1}{2} \int_0^1 dr r^2 p(r,t)$ with a fixed purification time $t$, and ask whether the locally optimal strategies formulated in the previous sections are \textit{globally} optimal for this goal. We note that in the ideal case (no decoherence and measurement efficiency $\eta=1$) global optimality of the locally optimal strategy (i.e., the unbiased measurement protocol when $\eta=1, \gamma_1=\gamma_\phi=0$) was proven in Ref. \cite{Wiseman:2008bc}. In the following subsections we investigate the extent of global optimality in other parameter regimes.

\subsection{No decoherence and $\eta\leq 1/2$}
\label{sec:TheEtaLeq12Case}

In this regime, we found the locally optimal control to be $u(r,t)\equiv \pm 1$, which is the diagonal measurement protocol. Here we use the verification theorem to prove that the locally optimal solution is actually globally optimal. Appendix \ref{sec:vt} reviews the verification theorem \cite{Jacobs:2008vt} and provides a simplified form that is useful for the present calculations.  The  verification theorem provides a sufficient set of criteria to test the global optimality of a presumed solution. We will use $P$ instead of $r$ as the state variable in this subsection to avoid complications deriving from the $r=0$ singularity in the dynamical equation for $r$ (see \erf{eqa:dynamics}).  At the first step, we need to calculate a cost function $C(P,t)$, which is defined as the average impurity $\langle L(T) \rangle =1-\langle P(T) \rangle $ at time $T$, given that the purity is $P$ at time $t$. For an arbitrary initial state, the no-feedback diagonal measurement protocol specifies an initial rotation of the qubit to the $z$-axis and then simple measurement in the $z$-basis (no feedback). Since the initial rotation is assumed to be instantaneous, the dynamics under this protocol is the subsequent motion along the $z$ axis. This density matrix evolution under measurement alone can be solved with the method of linear quantum trajectories \cite{Jacobs:2006lqt}. Given $z(0^+)=z_0$ (the $0^+$ time simply indicates the time after the instantaneous rotation to the $z$-axis), $z(t)$ can be written as:
\begin{equation}
    z(R(t))=\tanh(\mathrm{arctanh}(z_0)+\sqrt{2k\eta}R(t)),
\end{equation}
where $R(t)$ is a random variable whose distribution function at time $t$ is given by:
\begin{align}
    p(R,t)=&\exp(\frac{R^2}{2t}-k\eta t)\sqrt{\frac{1-z_0^2}{2\pi t}} \nonumber \\
	&\cdot\cosh(\mathrm{arctanh}(z_0)+\sqrt{2k\eta}R).
\end{align}

We find the cost function $C(P,t)$ using this distribution as
\begin{align}
    \label{eqa:cost}
 C(P,t)=&\frac{\exp[-k\eta(T-t)]\sqrt{2(1-P)}}{\sqrt{8\pi(T-t)}}\nonumber \\
 &\cdot \int^{+\infty}_{-\infty} \mathrm{sech}[\mathrm{arctanh}(\sqrt{2P-1})+\sqrt{2k\eta}R] \nonumber\\
 &\cdot \exp[-\frac{R^2}{2(T-t)}]\mathrm{d}R.
\end{align}

The $G$ function for our dynamics (\ref{eqa:puri_rate}), defined in appendix \ref{sec:vt}, is related to the derivatives of the cost function by:
\begin{align}
    \label{eqa:G_function}
   G(P,t)=&-k(2P-1)  \nonumber \\
	&\cdot [4\eta (P-1)^2 \frac{\partial^2 C}{\partial P^2}+(1-3\eta+2\eta P)\frac{\partial C}{\partial P}]u^2 \nonumber \\
	& -k(1+\eta -2P)\frac{\partial C}{\partial P}.
\end{align}

The derivatives $\frac{\partial C}{\partial t}$ and $\frac{\partial^2 C}{\partial P^2}$ are continuous over the interval $[0,T)$ as required by the verification theorem. The cost function in (\ref{eqa:cost}) gives the $G$ function a nonnegative coefficient in front of $u^2$ (including the minus sign) for all $P$ and $t$. The explicit form of this coefficient is derived in Appendix \ref{sec:coe}. Therefore, $u(P,t)\equiv \pm 1$ are the maximizers of the $G$ function, and the verification procedure concludes that the corresponding diagonal measurement protocol is globally optimal in this parameter regime.

\subsection{No decoherence and $1/2 < \eta < 1$}
\label{sec:The12Eta1Case}
In the regime $1/2 < \eta < 1$, the locally optimal strategy combines the diagonal measurement protocol and the unbiased measurement protocol, with a switch between these at a critical Bloch vector length $r^*$. In this case, however, one can show that this locally optimal strategy is not globally optimal. To do so, we can solve for average purity as a function of time and compare it against the corresponding purity derived from other protocols. In section \ref{sec:analytical_soln} we obtained an analytical form for the probability distribution for the Bloch vector length as a function of time, $p(r,t)$, when using the locally optial strategy. With this distribution function, the average purity, $\langle P(t) \rangle$, can be calculated by a simple integral.

\begin{figure}
	\centering
	\subfigure[The initial condition $r_0=0 < r^*$]{
  	\includegraphics[width=0.45\textwidth]{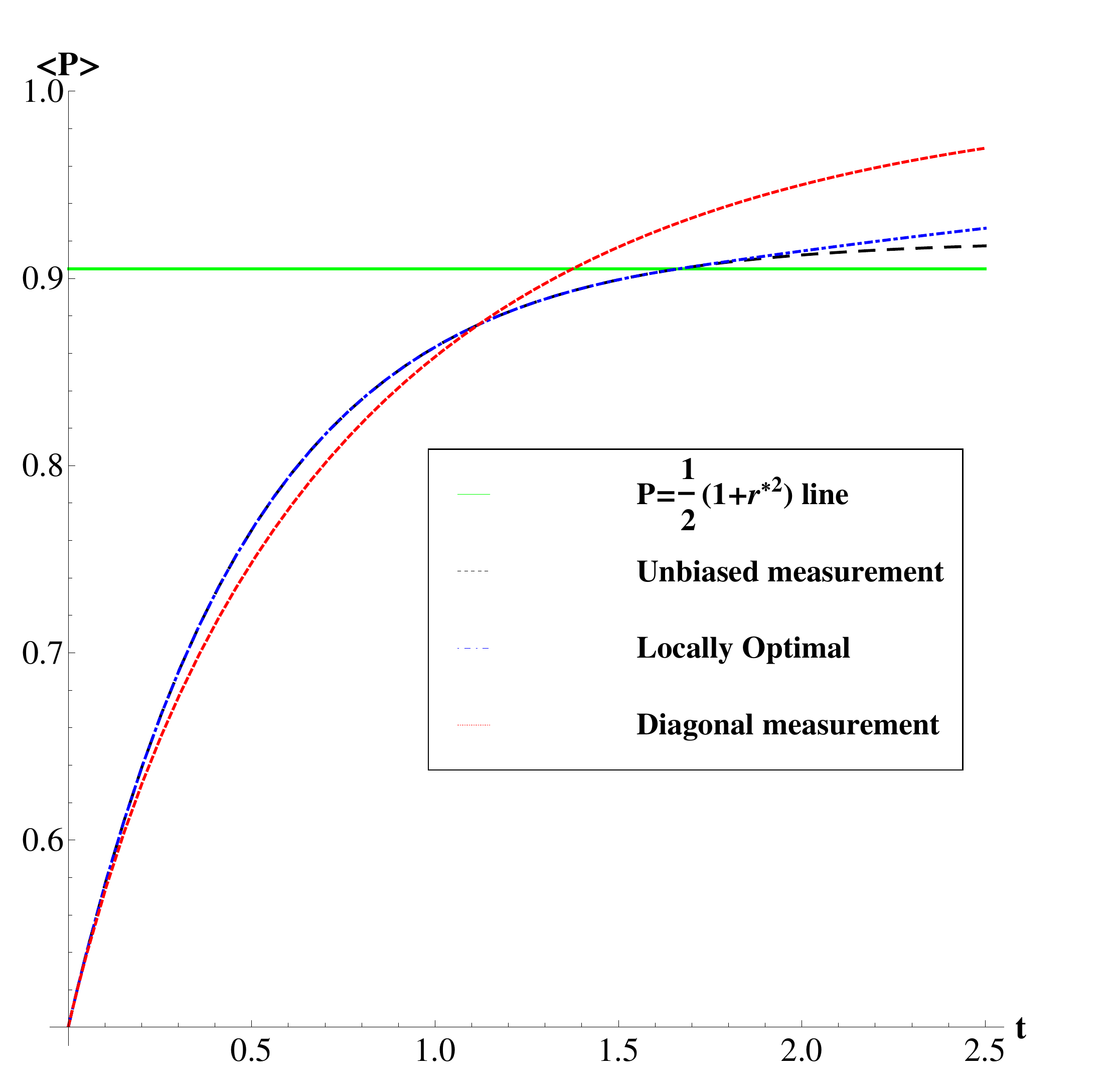}
   	\label{fig:noiseless_max_purity_1}
  }
  \qquad
	\subfigure[The initial condition $r_0=0.95 > r^*$]{
  	\includegraphics[width=0.45\textwidth]{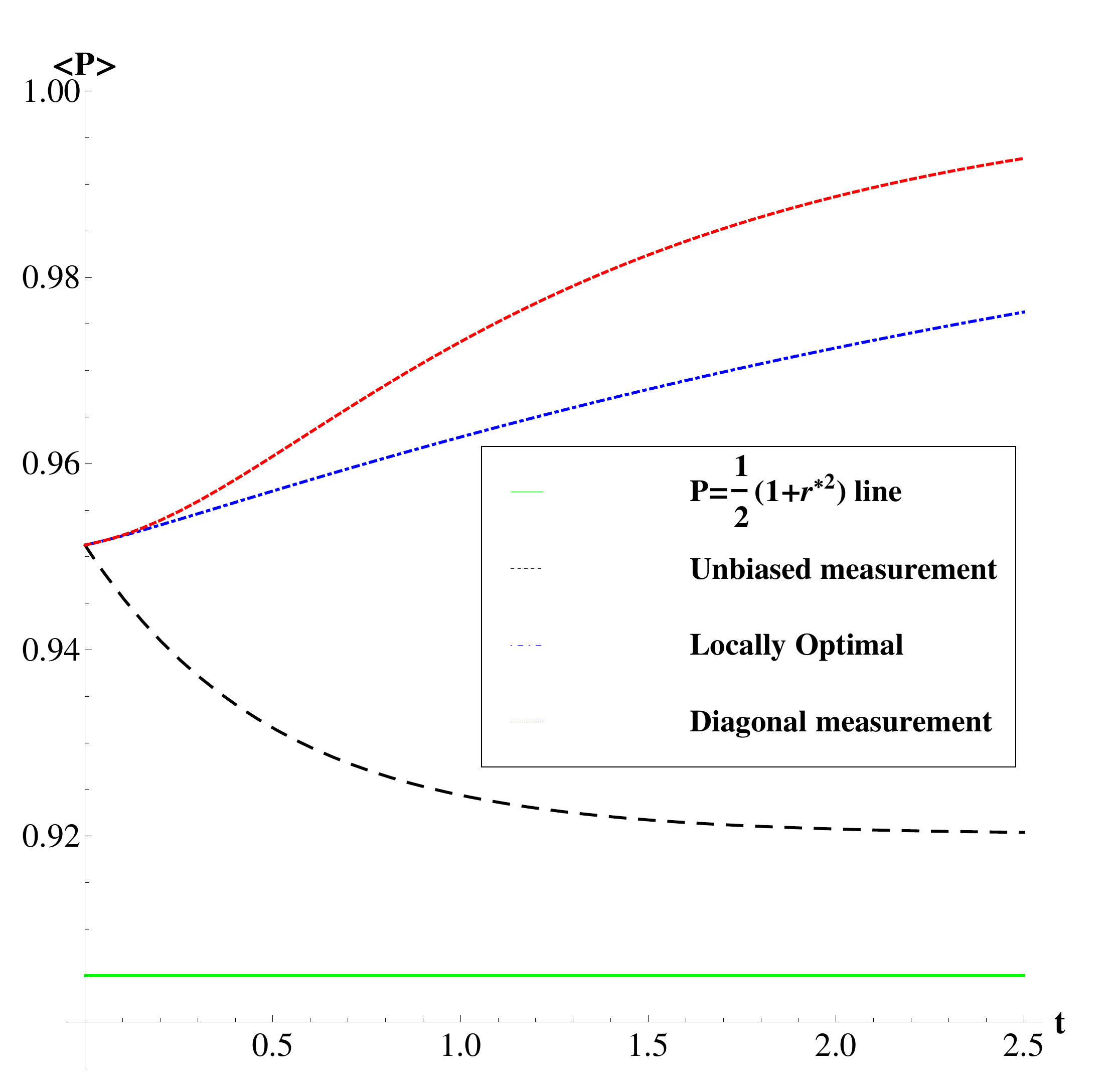}
   	\label{fig:noiseless_max_purity_2}
   }

   \caption{Average purity vs. time in the case of no decoherence for the unbiased measurement protocol, the locally optimal strategy, and the diagonal measurement protocol for two different initial purities. The parameters used here are $k=1$, $\eta=0.84$ ($r^*=0.9$), $\gamma_1=\gamma_\phi=0$.}
   \label{fig:noiseless_max_purity}
\end{figure}

Figures \ref{fig:noiseless_max_purity_1} and \ref{fig:noiseless_max_purity_2} show how the average purity evolves as a function of time (for three different control strategies) in the absence of decoherence, with parameters chosen so that $r_0<r^*$ and $r_0>r^*$, respectively. We have tested a wide range of parameter values and the results are qualitatively the same throughout this regime of measurement efficiency ($1/2 < \eta < 1$). 
With $r_0<r^*$, the locally optimal strategy initially outperforms the diagonal measurement protocol as expected. However, the diagonal measurement protocol (\textit{i.e.} constant $\sigma_z$ measurement) catches up later and purifies more effectively at late times. The catch up time occurs before the Bloch vector length reaches $r^*$ using the locally optimal strategy. 
When $r_0\geq r^*$, Fig. \ref{fig:noiseless_max_purity_2}, the locally optimal strategy never outperforms the diagonal measurement protocol \footnote{For very short times the difference between purities for the two protocols is within numerical error.}. 

\begin{figure*}
	\centering
	\subfigure[$t=0.05$]{
  	\includegraphics[width=0.45\textwidth]{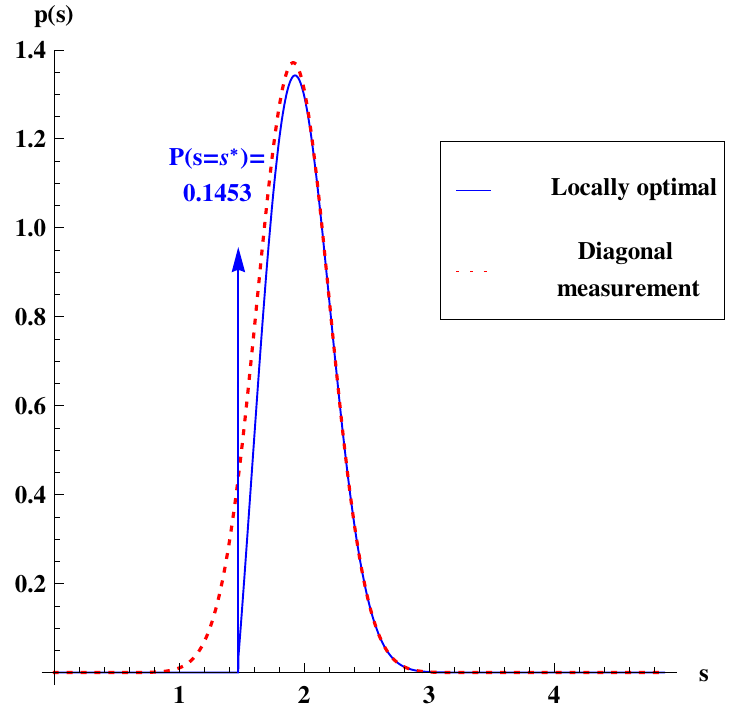}
   	\label{fig:noiseless_dist_1}
  }
  \qquad
	\subfigure[$t=0.5$]{
  	\includegraphics[width=0.45\textwidth]{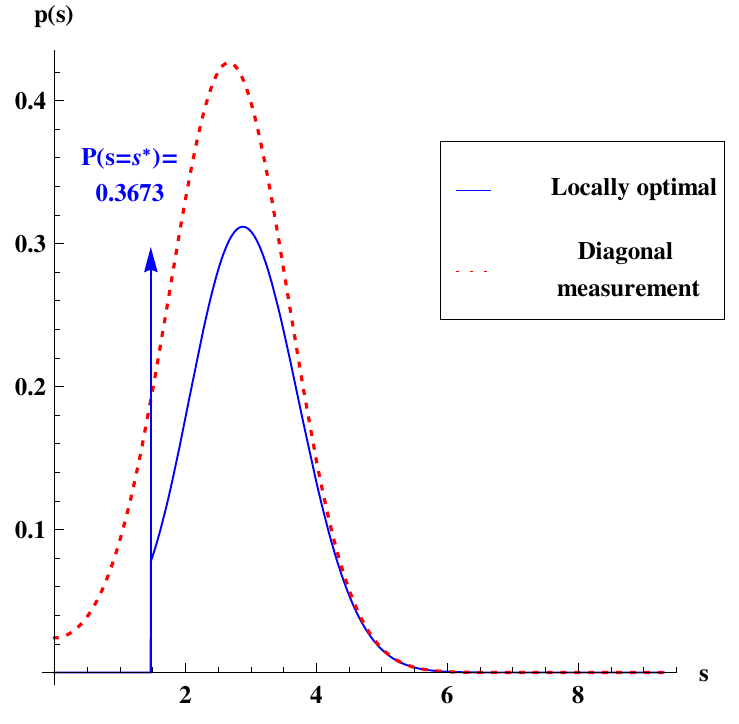}
   	\label{fig:noiseless_dist_2}
  }
  \qquad
	\subfigure[$t=5$]{
  	\includegraphics[width=0.45\textwidth]{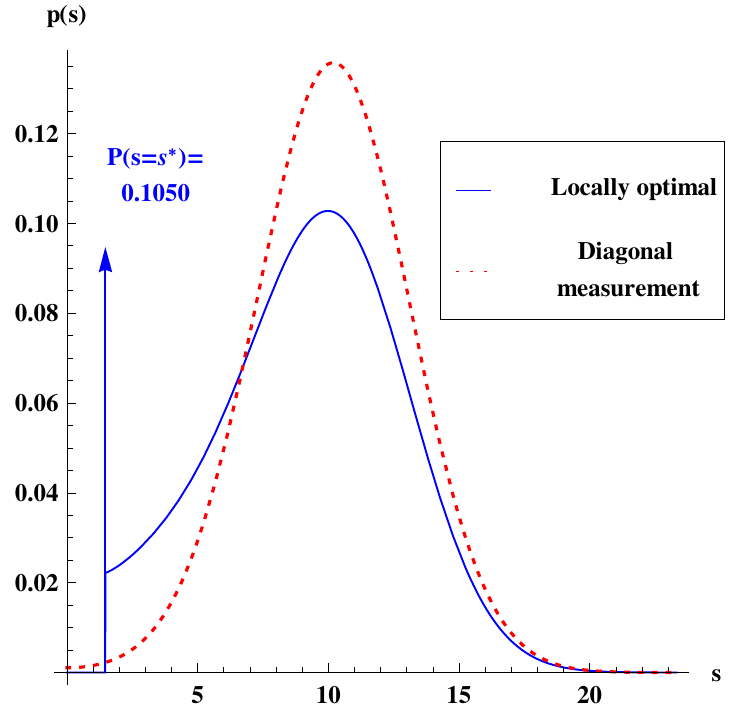}
   	\label{fig:noiseless_dist_3}
  }
  \caption{
Behavior of the solution to the Fokker-Planck equation, \erf{eqa:fokkerplanck} at three different times $t$, visualized by the distribution of $s(t)=\mathrm{arctanh}(r(t))$ as the qubit evolves under the locally optimal and the diagonal measurement protocols.  Both protocols operate on a qubit initially at the same purity, with $r_0=0.95$. Parameters used are the same as in Fig. \ref{fig:noiseless_max_purity}, namely $k=1$, $\eta=0.84$ ($r^*=0.9$), and $\gamma_1=\gamma_\phi=0$.}
  \label{fig:noiseless_dist}
\end{figure*}

The fact that there is a performance difference between the locally optimal strategy and the diagonal measurement protocol when $r_0 \geq r^*$ needs some explanation. When $r_0>r^*$, initially the locally optimal strategy is exactly the diagonal measurement protocol. However, as time progresses, while the diagonal measurement protocol simply causes diffusion of $r(t)$ along the $z$-axis, the locally optimal strategy switches to the unbiased measurement protocol if $r(t)$ drops below $r^*$ at any future time. The subsequent deterministic increase of purity caused by the unbiased measurement protocol results in $r(t+\Delta t)>r^*$ again for some small $\Delta t$. Then the switch back to the diagonal measurement protocol causes diffusion of $r(t)$ again. The net result of this switching back-and-forth at the boundary is a build-up of probability at $r=r^*$ at intermediate times. This concentration of probability weight at $r=r^*$ gives rise to a smaller cumulative probability in the $r>r^*$ region than the diagonal measurement protocol. This is illustrated in Figure \ref{fig:noiseless_dist}, which shows the time-development of the probability distribution, $p(r,t)$, when $r_0>r^*$ for both the locally optimal strategy and the diagonal measurement protocol. This probability concentration at the boundary is an interesting consequence of the switching behavior of the locally optimal strategy. This example demonstrates that while protocol switching can lead to local optimality, it can be detrimental to global optimality in some instances.

We also note that the unbiased measurement protocol eventually performs worse than both of the other protocols (diagonal measurement and the locally optimal strategy), regardless of the initial state. This demonstrates that using the optimal strategy derived for perfect efficiency measurements $\eta=1$ can be inappropriate if actually the measurement efficiency is less than unity.

From the above analysis we conclude that for the \texttt{max purity} goal, the locally optimal strategy, which maximizes the rate of increase of average purity at any time instant, is also globally optimal for both $\eta\leq 1/2$ and $\eta=1$. Therefore, feedback control is not helpful at all for the \texttt{max purity} goal when $\eta\leq 1/2$.  Instead,  the locally optimal strategy in this regime is simple measurement. In contrast, in the regime $1/2 < \eta <1$, we cannot conclude that any strategy is globally optimal.

\subsection{In the presence of decoherence}
\label{sec:max_purity_decoh}
\begin{figure*}
    \centering
    
    \subfigure[The locally optimal control]{
  		\includegraphics[width=0.43\textwidth]{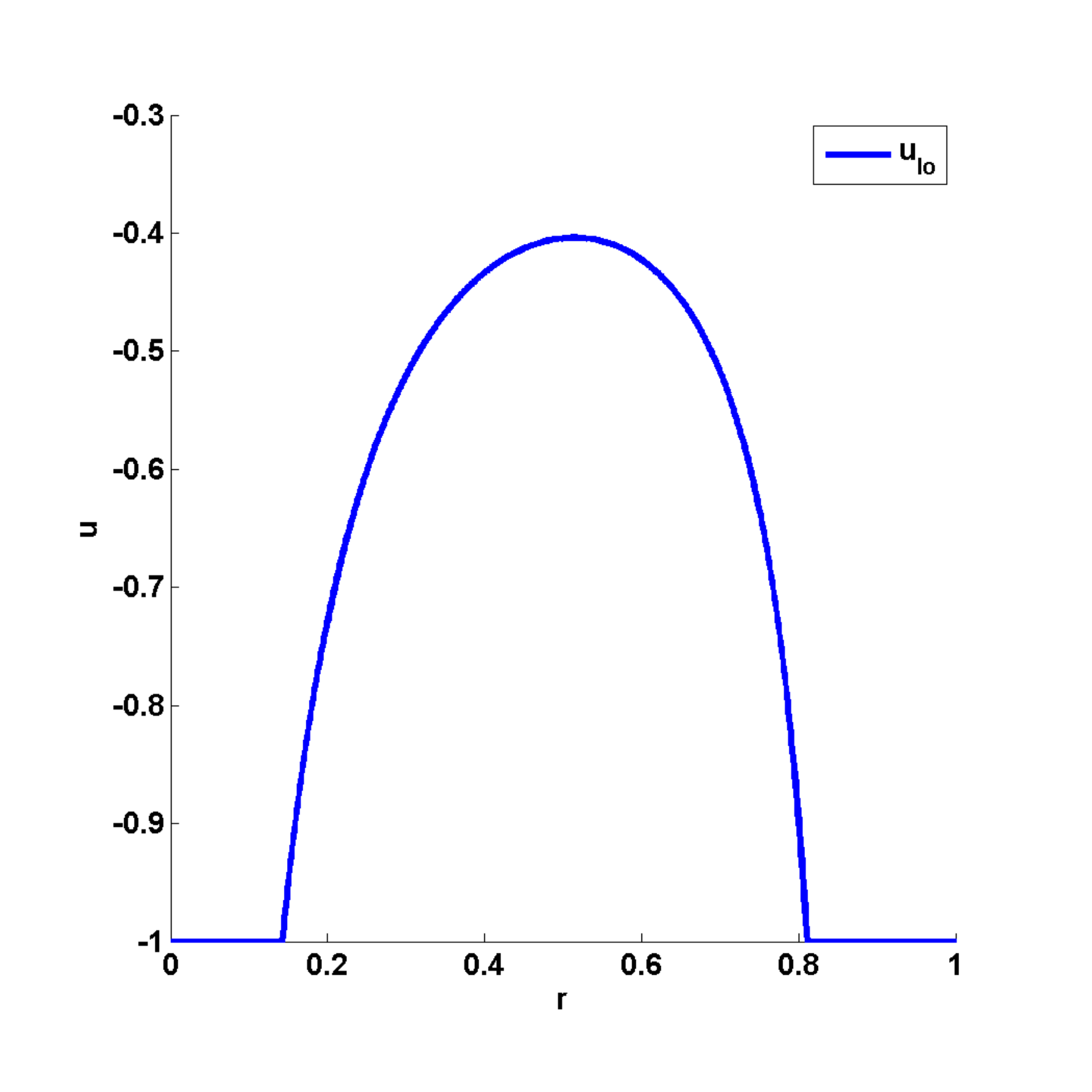}
   		\label{fig:noisy_u_lo}
  	}
  	
  	\subfigure[The average purity vs. time]{
  		\includegraphics[width=0.43\textwidth]{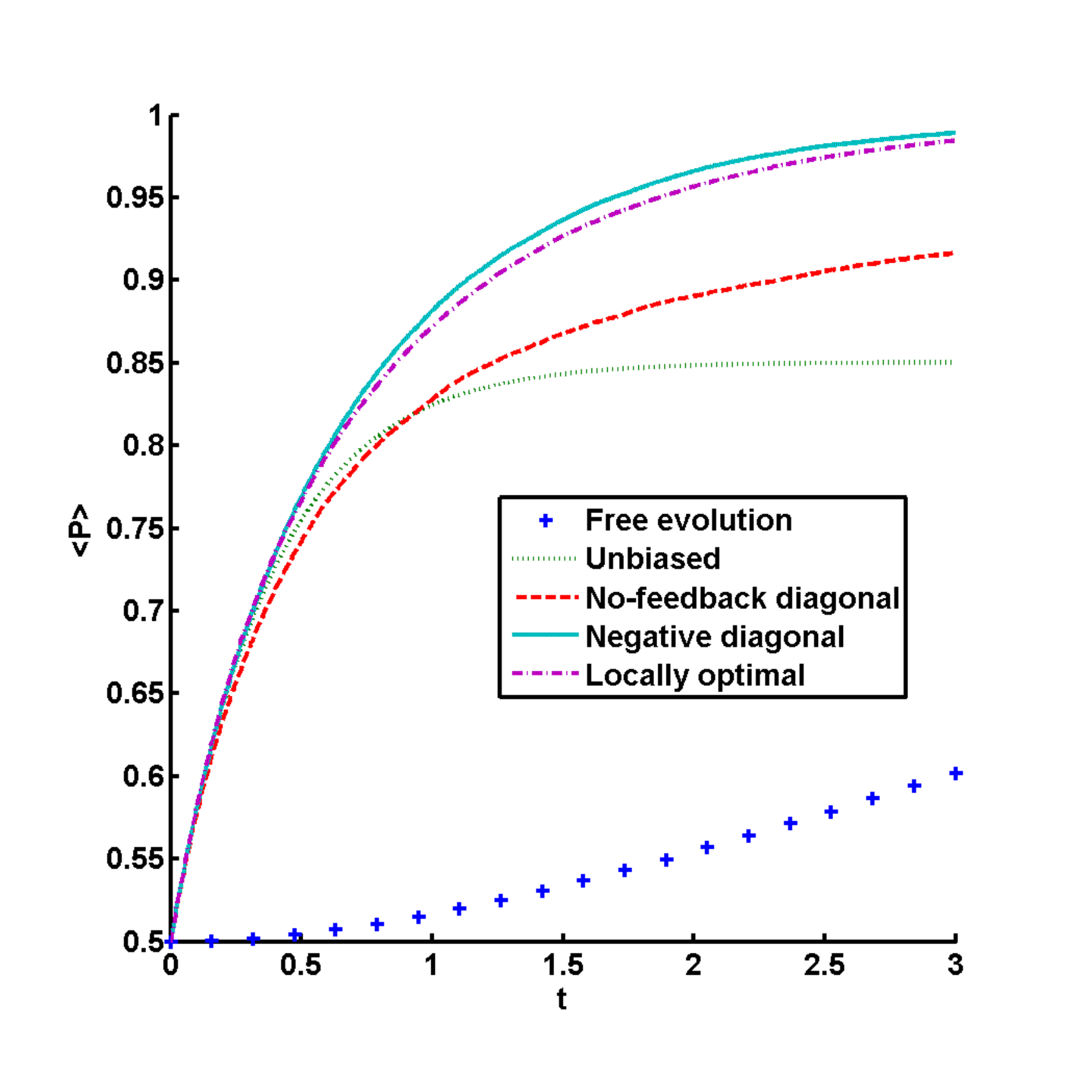}
   		\label{fig:noisy_max_purity}
  	}
  	\qquad
  	\subfigure[The average time vs. the target Bloch vector length]{
  		\includegraphics[width=0.43\textwidth]{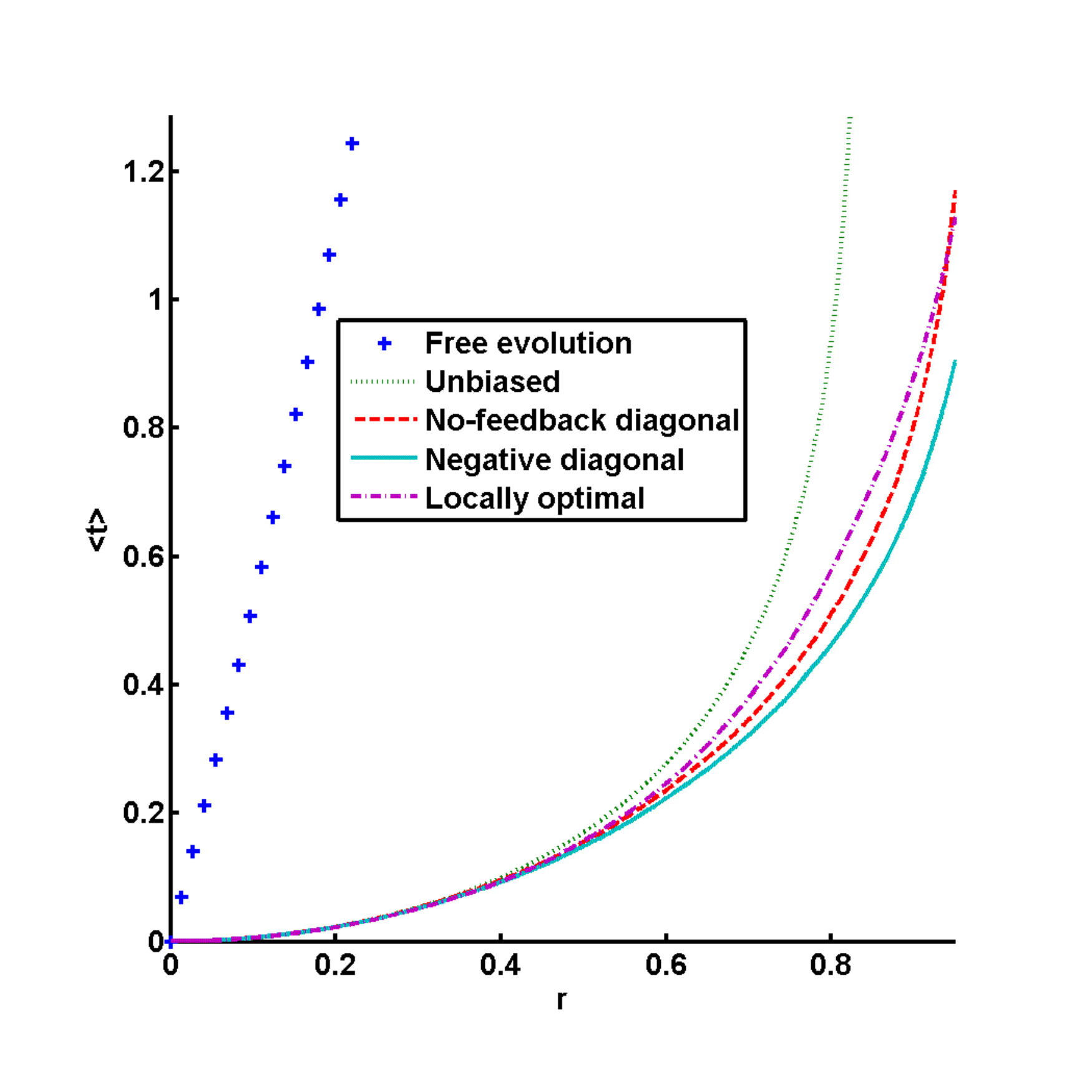}
   		\label{fig:noisy_min_time}
  	}
  	
    \caption{Performance comparison of the free evolution, the locally optimal strategy, the no-feedback diagonal measurement, the unbiased measurement, and the negative diagonal measurement protocols in the presence of decoherence for initial condition $\rho_0 = \mathbf{I}/2$ ($r_0=0$). The chosen parameters are $k=1$, $\eta=0.91$, $\gamma_1=0.2$, $\gamma_2=0.3$. \subref{fig:noisy_u_lo} plots the locally optimal strategy for this chosen set of parameters. Notice the switch from the preferred negative diagonal measurement control at small and large $r$ values ($u_{lo}(r,t)=-1$) to a more complex preferred control ($u_{lo}(r,t) > -1$) for intermediate values of $r$. \subref{fig:noisy_max_purity} shows the average purity (over 40,000 runs) vs. time and \subref{fig:noisy_min_time} shows the average purification time (over 20,000 runs) as a function of the target Bloch vector length for the five protocols. The statistical variation in the simulations is comparable to the line-width of the plots, so error bars are not explicitly shown.}
   
\end{figure*}

We have not been able to apply the verification theorem to prove global optimality of the locally optimal strategy for the \texttt{max purity} goal in any parameter regime in the presence of decoherence. However, we speculate that in the parameter regime where the negative diagonal measurement protocol is the locally optimal strategy (this regime is specified by the inequality $\gamma_1+2r[\gamma_2-\gamma_1+k(1-2\eta+\eta r^2)]\geq 0, \forall r$), that it is also the global optimum. This is because in this regime it is advantageous to cooperate with the relaxation process, which also induces purification, and this is precisely what the negative diagonal measurement protocol does. When the locally optimal strategy has a more complicated $r$ dependence (e.g., that in  Fig. \ref{fig:noisy_u_lo}), we numerically simulate the performance of the four different protocols we considered so far (using the Euler-Maruyama algorithm~\cite{Klo.Pla-1992} for those protocols involving stochastic integration). Fig. \ref{fig:noisy_max_purity} shows the behavior of average purity as a function of time for a typical set of parameters. The performance of the free evolution (with no measurement at all) is also added for comparison, since the relaxation dynamics itself induces purification. The negative diagonal measurement protocol is superior to the no-feedback diagonal measurement protocol as expected. It also achieves greater values of purity than the locally optimal strategy in the long run, similar to the situation found in the decoherence free case.
The locally optimal strategy  is also superior to the no-feedback diagonal measurement strategy \footnote{At least when the decoherence rates are not too small}.  The above facts suggest that, in the presence of decoherence, qubit purification measured by the \lowercase{\texttt{max purity}} goal will always benefit from feedback, even though in the absence of decoherence the corresponding preferred purification strategy may be no feedback (at long times).
 
\section{Global optimality for the \lowercase{\texttt{min time}} goal}
\label{sec:min_time}
In this section, we consider the \texttt{min time} goal, i.e., to minimize $-\int_1^0 t \mathrm{d}q(t)$, where $q(t)$ is the probability of not reaching a fixed target Bloch vector length at time $t$. We will determine whether the locally optimal strategies identified above are globally optimal in any parameter regime for this goal \footnote{We note that a convenient feature of the \texttt{min time} goal is that its evaluation does not depend on the definition of purity used; any definition of purity as an increasing function of $r$ is equivalent.}. 

\subsection{In the absence of decoherence}
\label{sec:min_time_no_decoh}
For the \texttt{min time} purification goal in the absence of decoherence, Wiseman and Bouten proved that the diagonal measurement protocol is globally optimal for perfectly efficient measurements,  $\eta=1$\cite{Wiseman:2008bc}.  We show here that in the absence of decoherence this is true for all values of measurement efficiency $\eta$ using the verification theorem. In this subsection, we will again use $P$ as the state variable, and $r$ should be interpreted as a function of $P$, namely, $r=\sqrt{2P-1}$.

Let $r_f$ be the final Bloch vector length that we want to achieve. The cost function $C(P,t)$ for this goal is the average remaining time of the first passage through $r_f$ ($r_f<1$), given that the qubit is at $r=\sqrt{2P-1}$ ($r\leq r_f$) at time $t$. We took the approach given in Ref. \cite{Wiseman:2006ti} to calculate this \texttt{min time} cost, and solved the Fokker-Planck equation with absorbing boundary conditions  at $r=r_f$. If $p(r,t)$ is the solution of this equation, then the probability of not hitting the target $r_f$ by time $t$ is $q(t)=\int_{-r_f}^{r_f} p(r,t)\mathrm{d}r$. The average time of reaching the boundary is then obtained from $q(t)$ by integration, namely $-\int_0^\infty t \dot{q}(t)\mathrm{d}t=\int_0^\infty q(t)\mathrm{d}t$. 

After the initial rotation to the $z$-axis that the diagonal measurement protocol prescribes, $z(0^+)=r$. Then $C(P,t)$ can be obtained as described above, and takes the form
\begin{equation}
    C(P,t)=(r_f\times\mathrm{arctanh}(r_f)-r\times\mathrm{arctanh}(r))/2k\eta.
\end{equation}
Substituting this form into the $G$ function, Eq. (\ref{eqa:G_function}), results in the following expression for the coefficient of $u^2$:
\begin{equation}
    1+\frac{1}{2}(\frac{r}{\eta}-\frac{1}{r})(\mathrm{arctanh}r+\frac{r}{1-r^2}),
    \label{eq:u2}
\end{equation}
where $r\in[0, r_f]$. 

It can be shown that Eq. (\ref{eq:u2}) is nonnegative if $\eta=1$. Since it is a decreasing function of $\eta$, the expression is also nonnegative for all $\eta$. The diagonal measurement protocol thus maximizes the function $G$, and by the verification theorem (appendix \ref{sec:vt}) it is globally optimal. Therefore in the absence of decoherence, feedback is not beneficial for the \texttt{min time} goal, regardless of the efficiency of measurement.

\subsection{In the presence of decoherence}
\label{sec:min_time_decoh}
In the presence of decoherence we cannot prove global optimality of any of the protocols, but we do have strong numerical evidence suggesting that the negative diagonal protocol is globally optimal for the \texttt{min time} goal in the presence of relaxation. 
We numerically simulated the qubit purification under the locally optimal, the no-feedback diagonal, unbiased measurement, and the negative diagonal measurement protocols with the Euler-Maruyama algorithm for many combinations of parameters. In Figure \ref{fig:noisy_min_time} we show results for one particular parameter combination, but qualitatively similar results are obtained for all parameter regimes simulated. We find that the negative diagonal measurement protocol is always superior to the other alternatives, particularly the locally optimal one and the no-feedback diagonal one. Therefore, we speculate that the negative diagonal measurement protocol constitutes the globally optimal strategy for the \texttt{min time} problem, regardless of decoherence rates or measurement efficiency. We also conclude that feedback is likely advantageous for the \texttt{min time} goal  in the presence of decoherence.

\section{Conclusions}
\label{sec:Conclusions}

Optimal feedback control can provide a crucial element of precision control in solid-state quantum systems, where the measurement process typically requires a non-negligible time to complete. In this work, we studied the optimality of feedback control protocols for qubit purification in the presence of decoherence and with realistic detectors characterized by non-ideal efficiency ($\eta <1$).  We considered the two control goals: (i) maximizing the average purity at a target time (\texttt{max purity}) and (ii) minimizing the average time to reach a specific purity threshold (\texttt{min time}).

When environmental decoherence is negligible and the only source of decoherence is the measurement back-action, we arrived at the following conclusions.  For the \texttt{max purity} goal and detector efficiency less than 1, we found that the globally optimal protocol is significantly different from the unbiased measurement protocol that is known to be globally optimal for this goal when $\eta=1$ \cite{Wiseman:2008bc}. This underscores the fact that one should be careful when extrapolating the optimality of feedback control protocols from idealized to realistic scenarios. The diagonal measurement protocol, which is an initial rotation to the $z$-axis followed by measurement in the $z$ basis, is optimal when $\eta \leq 1/2$, as analytically verified by the verification theorem. The diagonal measurement protocol also performs very well when $1/2<\eta<1$, where indeed it outperforms the locally optimal protocol in the long time limit.  We were however unable to find the global optimal solution in this regime.  In contrast, the situation for the \texttt{min time} purification goal  is quite different. Here the diagonal measurement protocol is known to be (globally) optimal solution for an ideal detector \cite{Wiseman:2008bc}.  In this work we showed that this optimality under the most ideal conditions holds for all values of detector efficiency $\eta$, as long as no decoherence is present.

We then explored the effects of decoherence on the optimal feedback strategies in addition to non-ideal ($\eta<1$) detectors. The decoherence sources were modeled by independent dephasing and relaxation processes under Markovian conditions. Here extensive numerical simulations show that the negative diagonal measurement protocol, which is designed to maintain the qubit in the negative segment of the $z$ axis by strong feedback, outperforms the other three feedback strategies (unbiased measurement, no-feedback diagonal, and locally optimal) for the \texttt{min time} goal, and for the \texttt{max purity} goal in the long time limit. Similar to the decoherence free case, there is a regime where the negative diagonal measurement protocol is not locally optimal. (It typically happens when the measurement efficiency is high and decoherence is weak.) Nevertheless, the negative diagonal measurement protocol achieves better average purity than the locally optimal one after certain time. The negative diagonal measurement protocol's good performance for both problems strongly suggests that feedback is useful in the presence of decoherence.

An interesting aspect of the study presented here is the behavior of the locally optimal protocol when $1/2<\eta<1$, and in the absence of decoherence. This protocol involves switching between two strategies, unbiased measurement and diagonal measurement, which corresponds to a switching between regions where the purity increases ballistically and diffusively, respectively. We demonstrated that this switching behavior results in a concentration of probability (of purity) around the boundary that defines the switching behavior. This novel aspect results directly from the dynamic switching of protocols and to the best of our knowledge has not been explored in other quantum control contexts. 

Overall, this work extends prior optimal control results in the quantum realm to include realistic experimental conditions and shows that significant modifications of optimal feedback control strategies can arise in the presence of decoherence.  In future it will be interesting to further extend these studies to analysis for qubits coupled to non-Markovian environments, to determine how the detailed behavior of an environment may enter the optimal control strategy.

\begin{acknowledgements}
The effort of HL, AR and KBW was supported by grants from NSA and DARPA. 
Sandia is a multi-program laboratory managed and operated by Sandia Corporation, a wholly owned subsidiary of Lockheed Martin Corporation, for the United States Department of Energy's National Nuclear Security Administration under contract DE-AC04-94AL8. 
\end{acknowledgements}

\appendix
\section{The Verification Theorem}
\label{sec:vt}
In this appendix, we review the procedure for verifying the (global) optimality of a given solution to a stochastic control problem.
An introduction to this topic can be found in Ref.\cite{Jacobs:2008vt}.
Consider the general dynamical equation for a stochastic system
\begin{equation}
   d\mathbf{x} = \mathbf{A}(t,\mathbf{x},\mathbf{u}(\mathbf{x},t)) dt + \mathbf{B}(t,\mathbf{x},\mathbf{u}(\mathbf{x},t)) \mathbf{dW} .
\label{eqa:sys-dy}
\end{equation}
Here the state of the system is given by the vector $\mathbf{x}$, and the vector $\mathbf{u}(\mathbf{x},t))$ denotes the control inputs. (The region that bounds $\mathbf{u}(\mathbf{x},t))$ shall not depend on $\mathbf{x}$ or $t$.) The vectors $\mathbf{A}$ and $\mathbf{B}$ are coefficients of the deterministic and stochastic parts of the dynamics, respectively.

The control objective is to minimize a cost, $J$:
\begin{equation}
  J =  \left\langle \int_0^T \!\! L(\mathbf{x},\mathbf{u}(\mathbf{x},s),s) ds + M(\mathbf{x}(T))  \right\rangle ,
\end{equation}
where $L(\mathbf{x},\mathbf{u}(\mathbf{x},t),t)$ is the cost rate, usually the consumed energy penalty, and $M(\mathbf{x}(T))$ is the cost of the final state at time $T$. 

The cost function, $C(\mathbf{x},t)$, is defined as the partial cost over the interval $[t,T]$, given that the system is at state $\mathbf{x}$ at time $t$:
\begin{equation}
  C(\mathbf{x},t) =   \left\langle \int_t^T \!\! L \; ds + M(\mathbf{x}(T))  \right\rangle . 
\end{equation} 

To determine whether a given control protocol, $\mathbf{u_c}(\mathbf{x},t)$ is optimal, one performs the following three steps:

\textbf{1}. Integrate the equations of motion of the system to calculate the cost function, $C(\mathbf{x},t)$, for this protocol.

\textbf{2}. Check that $C$ satisfies two continuity conditions:
\begin{equation}
        \frac{\partial C}{\partial t}  \;\;\;\;\; \mbox{and} \;\;\;\;\;  \frac{\partial^2 C}{\partial \mathbf{x}^2}
\end{equation}
are continuous. Here $\partial^2 C/\partial \mathbf{x}^2$ denotes the matrix of second derivatives of $C$. 

\textbf{3}. Determine whether or not $\mathbf{v}(\mathbf{x},t)=\mathbf{u_c}(\mathbf{x},t)$ is a maximizer of the following function of $\mathbf{v}$:

\begin{align}
  G(t,\mathbf{x},\mathbf{v}) =& - \frac{1}{2} \mbox{Tr} \left[ \mathbf{B}^{\dagger}(t,\mathbf{x},\mathbf{v}) \frac{\partial^2 C}{\partial \mathbf{x}^2} \mathbf{B}(t,\mathbf{x},\mathbf{v})  \right] \nonumber\\
	&-  \mathbf{A} \boldsymbol{\cdot} \frac{\partial C}{\partial \mathbf{x}}- L(t,\mathbf{x},\mathbf{v}) .
\label{eqa:G}
\end{align}

Note that one must check that $\mathbf{u_c}(\mathbf{x},t)$ maximizes $G$ separately at each time $t$ and at each value of $\mathbf{x}$.

The above verification procedure provides a sufficient condition for a control strategy to be optimal. In Ref.\cite{Jacobs:2008vt}, the procedure has four steps, and we have removed the third step by realizing that the  Hamilton-Jacobi-Bellman equation is automatically satisfied if the $\mathbf{u_c}$ maximizes Eq. (\ref{eqa:G}).

For a time-optimal control problem where the goal is to minimizing the average time taken for a function  $h(\mathbf{x}(t),t)$ of the dynamical variables (and perhaps of time) to cross a fixed threshold $h_c$, the same three-step verification procedure still holds. Please note that the cost function $C(\mathbf{x},t)$ should be defined as the average remaining time it will take to cross the threshold,  given that the current time is $t$ and current state is $\mathbf{x}$. The corresponding function $G$ in step 3 is defined as

\begin{equation}
  G(t,\mathbf{x},\mathbf{v}) = - \frac{1}{2} \mbox{Tr} \left[ \mathbf{B}^{\dagger}(t,\mathbf{x},\mathbf{v}) \frac{\partial^2 C}{\partial \mathbf{x}^2} \mathbf{B}(t,\mathbf{x},\mathbf{v})  \right] -  \mathbf{A} \boldsymbol{\cdot} \frac{\partial C}{\partial \mathbf{x}}.
\end{equation}

\section{The Coefficient of the $u^2$ Term}
\label{sec:coe}

In this appendix, we will prove that the coefficient of the $u^2$ term in Equation \ref{eqa:G_function} is nonnegative for all $P\in[\frac{1}{2}, 1]$, $\eta\in[0, 0.5]$, $k>0$, and $T-t>0$ with $C(r,t)$ given by Equation \ref{eqa:cost}.

The coefficient works out to be the following expression up to a positive factor:
\begin{align}
\label{eqa:coefficient}
& \int^{+\infty}_{-\infty}\big\{ 2\eta r(1-r^2) \mathrm{sech}^2[\mathrm{arctanh}(r)+\sqrt{2k\eta}R] \nonumber\\
&+ (r^2-\eta)\{r+\tanh[\mathrm{arctanh}(r)+\sqrt{2k\eta}R]\} \big\} \nonumber \\
&   \cdot \mathrm{sech}[\mathrm{arctanh}(r)+\sqrt{2k\eta}R]\exp(-\frac{R^2}{2(T-t)})\mathrm{d}R .
\end{align}

First, we ignore $\eta$ in the expression $k\eta$ because it can always be absorbed by $k$, and then we divide the above expression by $\eta$. It is easy to see that the resultant expression is a decreasing function of $\eta$, because the following integral is nonnegative from symmetry analysis:
\begin{align}
 &   \int^{+\infty}_{-\infty}\tanh[\mathrm{arctanh}(r)+\sqrt{2k\eta}R]  \nonumber \\
  &  \cdot\mathrm{sech}[\mathrm{arctanh}(r)+\sqrt{2k\eta}R]\exp(-\frac{R^2}{2(T-t)})\mathrm{d}R.
\end{align}

Therefore, we only need to prove the positivity of Expression \ref{eqa:coefficient} for $\eta=\frac{1}{2}$. Expanding the hyperbolic functions with the addition formulas yields the following expression for the integrand:
\begin{align}
&    \frac{2r^5+(2r^2-1)(3r^2+1)\tanh R+r(5r^2-3)\tanh^2 R}{2(1+r\tanh R)^3}\nonumber\\
&   \cdot \sqrt{1-r^2}\mathrm{sech}R\exp(-\frac{R^2}{2(T-t)}).
\end{align}

Because an odd function of $R$ contribute nothing to the integral, we only need to take into account the even component of the integrand, which turns out to be the following expression up to a positive factor:
\begin{align}
&    (-6r^7+16r^5-8r^3)\tanh^4 R \nonumber\\
&   +(6r^7-18r^5+8r^3)\tanh^2 R +2r^5.
\end{align}

From the properties of parabolic functions, it can be shown that the above expression, and thus the even component of the integrand, is nonnegative for all $R$. Therefore, Expression \ref{eqa:coefficient} is nonnegative.

\section{Explicit Solution of Equation (\ref{eqa:pde})}
\label{sec:solvingpde}

Eqs. (\ref{eqa:pde}a) and (\ref{eqa:pde}b) can be combined to give a boundary condition for $p_2(r,t)$. Then we would like to make a change of variables. Let $r=\tanh s$ (similarly for $r^*$ and $r_0$), and let $p_2(r,t)=\mathrm{sech}s_0\mathrm{e}^{-k\eta t}\cosh^3s Q(s,t)$. Eq. (\ref{eqa:pde}) gets translated into the following equations with the initial condition $Q(s,0)=\delta(s-s_0)$.
\begin{subequations}
    \label{eqa:heat}
    \begin{align}
        & [- Q(s,t) - A\frac{\partial}{\partial s}Q(s,t) +B\frac{\partial}{\partial t}Q(s,t)]|_{s=s^*}=0,\\
        & \frac{\partial}{\partial t}Q(s,t)=k\eta\frac{\partial^2}{\partial s^2}Q(s,t), s>s^*,
    \end{align}
\end{subequations}
where $A=\mathrm{csch}^2 s^* \coth{s^*}$, $B=\coth^2 s^*/(k\eta)$.

If $Q(s,t)$ is a solution to the above equations, it is easy to see that $q(s,t)=- Q(s,t) - A\frac{\partial}{\partial s}Q(s,t) +B\frac{\partial}{\partial t}Q(s,t)$ is a solution to the same heat equation (\ref{eqa:heat}b) with the boundary condition (\ref{eqa:heat}a) replaced by $q(s^*,t)=0$. We can express $Q$ in terms of $q$ as:
\begin{align}
    \label{eqa:Q}
    Q(s,t)=&-\frac{1}{A}\int^{0}_{-\infty}\exp(\frac{x}{A})q(s+x,t-\frac{Bx}{A})\mathrm{d}x\nonumber\\
&+\exp(-\frac{s}{A})f(\frac{B}{A}s+t),
\end{align}
where $f(x)$ is an arbitrary function of $x$.

The solution to the heat equation with no boundary is given by:
\begin{equation}
    Q_0(s,t)=\frac{1}{2\sqrt{\pi}\eta kt}\exp(-\frac{(s-s_0)^2}{4\eta kt}).
\end{equation}

Let $q_0(s,t)=- Q_0(s,t) - A\frac{\partial}{\partial s}Q_0(s,t) +B\frac{\partial}{\partial t}Q_0(s,t)$. We set $q(s,t)=q_0(s,t)-q_0(2s^*-s,t)$ so it satisfies the heat equation and vanishes at $s^*$. In order for $Q(s,t)$ in Eq. (\ref{eqa:Q}) to meet the initial condition, we also need to set $f(x)=D\exp(\frac{AC+1}{B}x)$, where:
\begin{subequations}
   
    \begin{align}
        & C=\frac{A-\sqrt{A^2+4B\eta k}}{2B\eta k},\\
        & D=\frac{2AC\exp[C(s_0-2s^*)]}{\sqrt{A^2+4B\eta k}}.
    \end{align}
\end{subequations}

It can be verified that the resulting $Q(s,t)$ given by Eq. (\ref{eqa:Q}) is the desired solution.


\bibliography{purification}

\clearpage
\appendix
\end{document}